\documentclass[%
 aip,
 amsmath,amssymb,
 reprint,%
]{revtex4-1}

\usepackage{graphicx}
\usepackage{dcolumn}
\usepackage{bm}

\usepackage[utf8]{inputenc}
\usepackage[T1]{fontenc}
\usepackage{mathptmx}
\usepackage{etoolbox}
\usepackage{amsmath}
\usepackage{mathtools}
\usepackage{multirow}
\usepackage{bigints}

\makeatletter
\def\@email#1#2{%
 \endgroup
 \patchcmd{\titleblock@produce}
  {\frontmatter@RRAPformat}
  {\frontmatter@RRAPformat{\produce@RRAP{*#1\href{mailto:#2}{#2}}}\frontmatter@RRAPformat}
  {}{}
}%
\makeatother
\begin{document}

\preprint{AIP/123-QED}

\title{How soap bubbles change shape while maintaining a fixed volume of air?}

\author{Wei-Chih Li$^{1,2}$, Chih-Yao Shih$^{1}$, Tzu-Liang Chang$^{1}$, and Tzay-Ming Hong }
\thanks{ming@phys.nthu.edu.tw}
\affiliation{$^1$Department of Physics, National Tsing Hua University, Hsinchu, Taiwan 30013, Republic of China\\$^2$Department of Physics, Emory University,
Atlanta, Georgia 30322, USA}\

\date{\today}

\begin{abstract}
We combine experiments and theoretical derivations to study the evolution of a stretched soap bubble and compare it with an open film to highlight the effect of volume conservation. There exists a critical length for both surfaces beyond which the bottleneck developed in their middle starts to shrink irreversibly and pinch off into multiple compartments. Before the system leaves the regime of equilibrium, the minimization of surface energy plays a major role in determining its shape, which can be tackled by the variational method theoretically. In contrast to open films, soap bubble volume conservation introduces a Lagrange multiplier that plays the role of pressure difference in mediating the evolution of bubble shape over a long range. By examining how boundary constraints influence bubble deformation, we establish a framework contrasting the bubble’s convex-to-concave transition with the behavior of soap films under similar conditions. Using experiments and theoretical modeling, we analyze the equilibrium and breakup regimes, providing insight into the role of geometric constraints on surface tension-driven systems. Our findings reveal critical differences between bubble and film stability profiles, shedding light on universal behaviors in non-equilibrium fluid mechanics and potential applications across biological and material sciences. 
\end{abstract}

\maketitle

\section{\label{sec:level1}Introduction}

The study of how droplets change their surfaces under different boundary conditions and what shape they adopt to minimize their surface energy while maintaining a fixed internal volume has deep historical roots in the field of variational calculus \cite{minimal_surface_Euler_Leonhard, goldstine1980history}. The concept can be traced back to the 18th century, prominently featured in the works of Euler and Lagrange who laid the foundational principles. Without the constraint of volume conservation, the problem is simple with the first successful example by Plateau \cite{Plateau} on the profile of open soap films.


The role of surface tension, or equivalently minimization of surface in liquid, is crucial in many fields.
Applications include (1) Biology: the structure of cell membranes that directly affects the activity of membrane-bound proteins. Building models for the dynamics of membranes is a crucial step in finding the mechanism behind the formation of the membrane structure. All of these models require minimizing the surface energy to stabilize the membrane structure, especially during the phase transition. 
(2) Chemical engineering: Help predict the structures and behaviors of colloidal membranes composed of rod-like viruses. Modifying the boundary conditions can change their shape from a catenoid to tethers under external forces. (3) Material science: For metal crystals in zeolites and other porous materials, surface tension dictates the arrangement of atoms to stabilize the structure. 

However, pure academic studies include the recent highlight of the fundamental role of catenoids and helicoids for equilibrium systems. Dimensional reduction to a one-dimensional Schrödinger operator has simplified the analysis of unstable modes \cite{minimal_surfaces_and_soap_film_instabilities}. Additionally, it has been shown that all minimal surfaces can be constructed from pieces of helicoids and catenoids \cite{Shapes_of_embedded_minimal_surfaces}. Combining theoretical proofs with numerical methods, the existence and properties of the genus-one helicoid have also been established by assuming periodic boundary conditions and employing the intermediate value theorem \cite{An_embedded_genus-one_helicoid}. 

Blowing soap bubbles is a childhood experience shared by most people. When the bubbles are stretched to some critical length, the bottleneck developed at their middle starts to shrink spontaneously. A notable outcome of this out-of-equilibrium process is
the pinch-off when the fluid breaks into smaller compartments, which is of practical importance in industrial applications such as inkjet printing \cite{material_printing} and injection molding \cite{molding}. It can even offer insight into biological processes, such as animal cell division \cite{cell}. 
Since this study focuses on long-range mediation of internal pressure due to volume conservation, we will concentrate on systems for which the surface tension dominates the viscous stress in influencing the dynamics of such a singularity. Examples include thin films \cite{PhysRevE.104.035105, yc}, droplets \cite{Hydrodynamic_stability, Lister_H_stone, slender, bir_bridge, Universal_Eggers, bir_pendant_drop, Iterated_Instabilities_nagel, Drop_formation_eaggers, inital_nagel, droplets_from_liquid_jet, Drop_formation_in_viscous_flows, Computational_experimental_drop_formation, Satellite_drops_pinch_off, Drop_formation_from_a_capillary_tube_one_d, Modeling_pinchoff_and_reconnection, Simplicity_and_complexity_in_dripping, Computational_analysis_of_drop, pinch_rewiew, Pinching_Dynamics_and_Satellite_Droplet_Formation_in_Symmetrical_Droplet_Collisions, two_fluid_snap, self_differ_viscosity, Testing_for_scaling, self_viscosity_thm, Computational_analysis_pinch_off_sel_similar, oscillating_pat_self_similarity, superfluid}, air bubbles \cite{Breaking_of_liquid_films_and_threads, giant_bubble_pinch}, and ligaments \cite{Ligament_Mediated_Spray_Formation} where universal power laws are known to dictate the inviscid fluid behavior. 
Numerical and theoretical studies \cite{Chen_steen_numerical, Self_Similar_Capillary} have confirmed the universality of these power laws. For example, Chen and Steen \cite{Chen_steen_numerical} demonstrated a power-law fit of $h_{\rm{min}}(\tau)$ with an exponent of $2/3$ during pinch-off. 
This prediction was later experimentally validated by Robinson and Steen \cite{robinson_steen_exp, cryer_steen_exp}, although they fell short of exploring the final behavior right before the breakup due to the limited frame rate of their high-speed camera.

The minimization of surface energy also plays a significant role in the evolution of systems at non-equilibrium. For instance, in axisymmetric membranes with areas exceeding the critical threshold, additional cylindrical tether solutions appear, 
making it possible to remain in equilibrium at increasingly large ring separations \cite{buckling_minimal_surfaces_tethers}. 
Understanding fluid behaviors in the pre-pinch-off regime, where instabilities in the catenoidal shape lead to a universal cone angle, is a crucial point focused in previous studies \cite{PhysRevE.104.035105}. 
During the collapse of catenoidal soap films, a unique geometric transition occurs, where two acute-angle cones are formed and connected to the supporting rings, joined by a central quasi-cylindrical region before the final pinch-off event.

The stability of soap membranes plays a crucial role in the study of two-dimensional flow. Traditionally, the lifespan of these membranes is prolonged by reducing the evaporation rate of the soap water, achievable by adding glycerol \cite{glycerol}. Frazier {\it et al.} \cite{big_bubble} have provided a quantified analysis of the role of long-chain polymers in the solution. Their findings bridge the gap between the effects of extensional rheology and the longevity of the membrane. In the meantime, Pasquet {\it et al.}\cite{detailed_recipe} offers a more detailed recipe, indicating the optimal ratio of surfactants, lubricants, and glycerol.

The breakage of liquid stored between the gap of two horizontally and vertically separated rods has been studied\cite{Simulation_ink_gravure, Stability_of_liquid_bridges_between_equal_disks_in_an_axial_gravity_field, vertical_abtrary_liquid_bridge, On_the_breakup_of_viscous_liquid_threads, nonlinear_bridge}. It was observed that the volume of the liquid influenced $L^*$; however, their conclusions were affected by the movement of the contact line \cite{contact_line} during neck collapse and the deviation from gravity in horizontal and vertical experiments.

The appeal of soap bubbles as a research subject lies in their simplicity and accessibility. Following this principle, the theoretical studies of our work are 
accompanied and backed up by experiments. Therefore, it is necessary to mention two technical details. 
First, the artifacts from gravity can be neglected by stretching a soap bubble horizontally. 
This is the place a third way of approach enters, i.e., we numerically calculate the profile of the soap bubble to show that the deviation due to gravity is indeed negligible because of the thin membrane of the bubble. 
Second, both the soap bubble and its boundaries at both ends are hollow. The motion of the soap water on the boundary is limited by the thickness of the hollow boundary caps. 

\begin{figure*}
	\centering	
	\includegraphics[width=18cm]{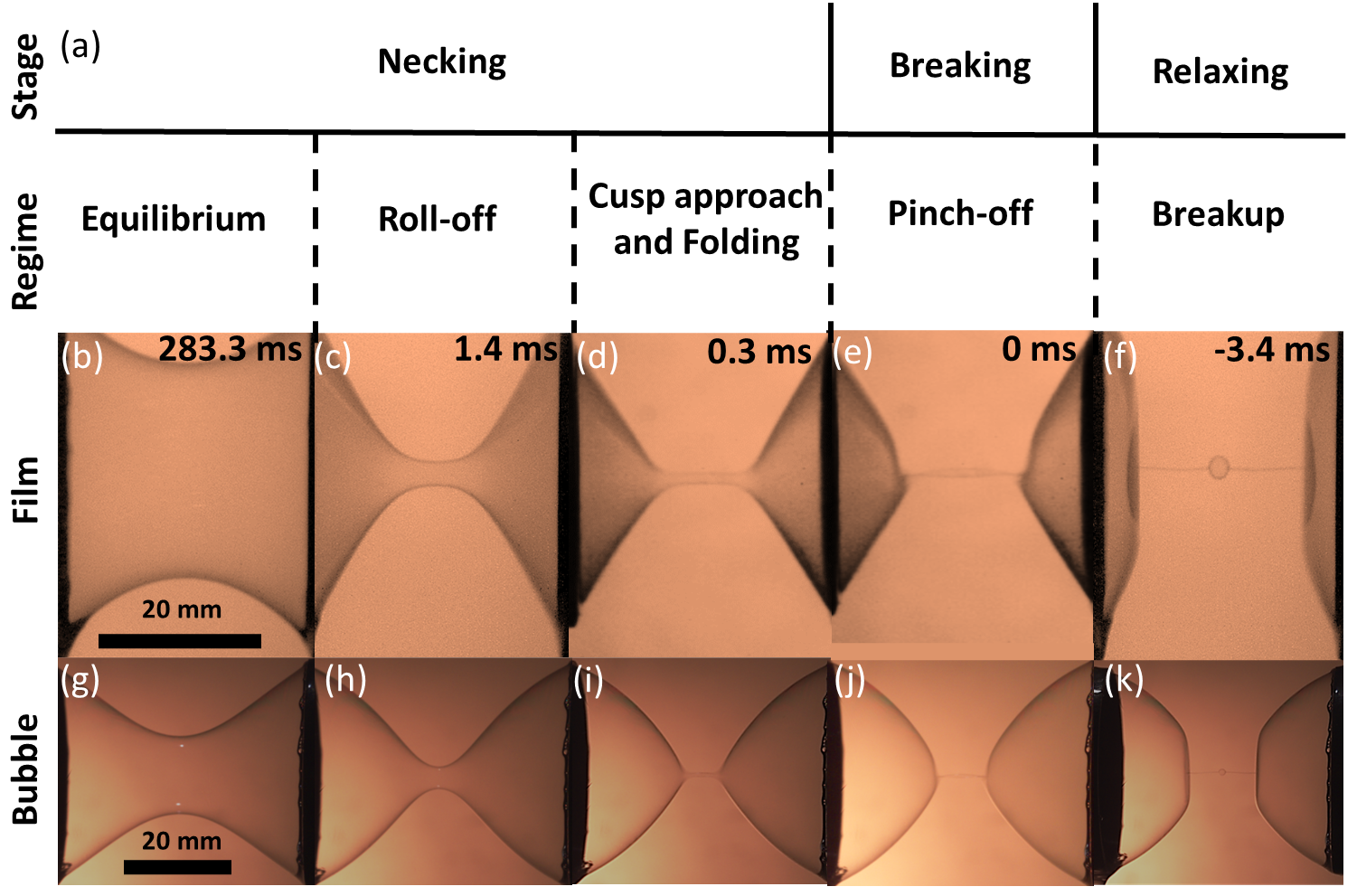}
	
    \caption{(a) The procedures of breakage consist of three stages which can be further divided into five regimes. Their corresponding photos at different $\tau$ for
  the soap film and bubble are shown in (b$\sim$f) and (g$\sim$k), separately.  Note that the film is allowed to squeeze air out of its interior, while the bubble has to roughly conserve its volume. The pulling speed is set at $v_{s}=$ 16 mm/s, the radius of ring or cap $R=$ 20 mm, and the pumping volume of bubble $V=$ 26 ml. }
    \label{fig:bubble_set}		    
\end{figure*}

\section{Experimental setup}
\begin{figure*}
	\centering
	\includegraphics[width=13cm]{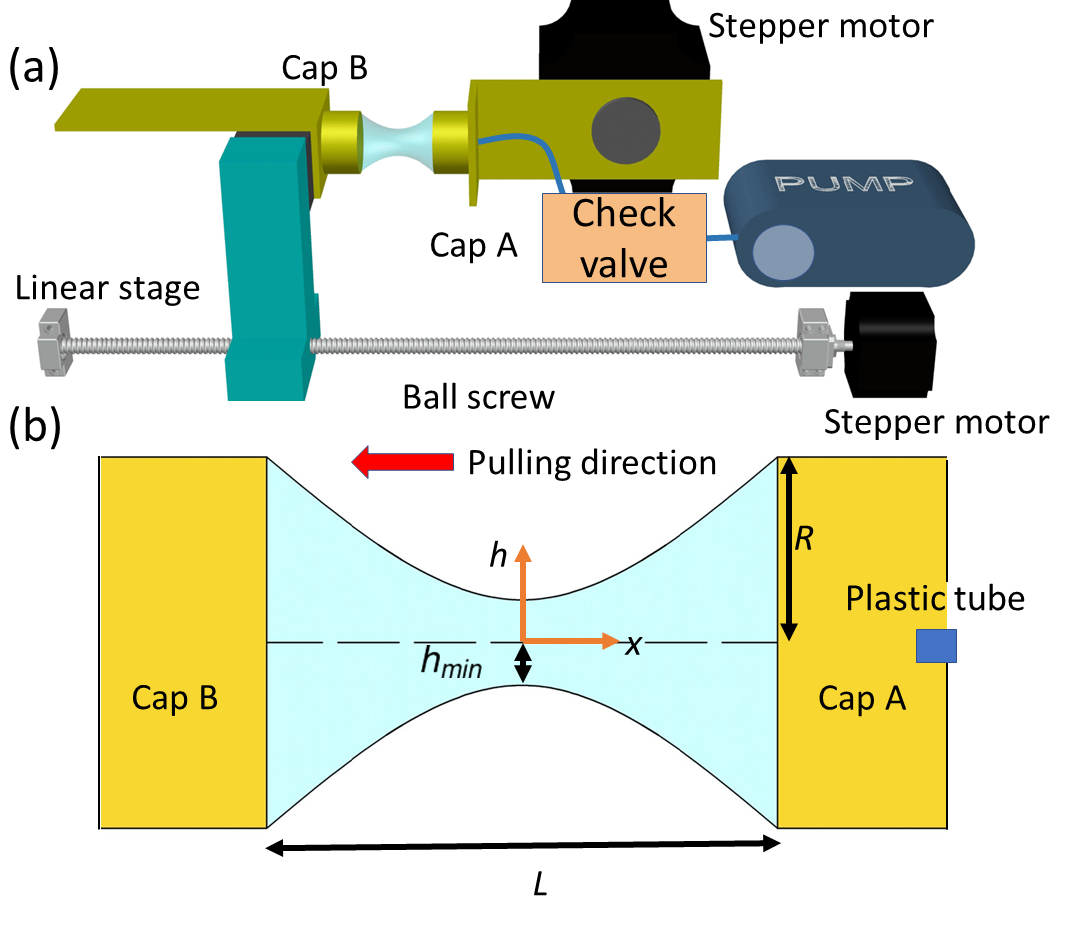}
    \caption{(a) Schematic experimental setup for stretching soap bubble by a stepper motor. (b) Relevant parameters are defined. The bubble is painted in blue, while the caps are in yellow. Cap A and B are replaced by Ring A and B to produce a film.}
    \label{fig:set_flat}
\end{figure*}

Our soap water contains the following ingredients: soap made from dried oleic acid, filtered water, and powdered guar gum. The addition of guar gum has been verified to prolong the lifetime of the soap membrane \cite{big_bubble}. Following immersion in the solution, an aluminum cap A with a radius of $R$ is rotated by 90 degrees using a stepper motor to horizontally align its open end with cap B, which is positioned at a distance of $L$. When the air pump is activated using a solid-state relay module, a soap bubble is generated on cap A. To prevent deflation of the bubble, we incorporate a check valve to ensure that no air backflow occurs. As demonstrated schematically in Fig.   \ref{fig:set_flat}, this bubble is gently attached to cap B that is pre-wetted. Next, cap B is moved away from cap A using a linear ball screw driven by another stepper motor. The pulling speed is maintained at a constant value of approximately $v_{s} \approx 16$ mm/s. The collapse of the bubble neck is captured by a high-speed camera operating at 23000 fps. 
It is worth mentioning that the soap film was originally prepared by drilling a big hole in cap B to allow the air to flow freely out of the membrane. However, it turned out that the close surface of cap A introduced asymmetry to the airflow that consequently distorts the shape of the soap film. Finally, we did away with the cap design altogether and used open rings. 

After several trial-and-errors, we ultimately decided on the following ingredients for our soap water: soap made from dried oleic acid, filtered water, and powdered guar gum.  The recipe can be found in Table \ref{recipe}.
The viscosity of soap water is measured by a Cannon-Fenske viscometer which works by recording the time it takes for fluid to flow through a narrow tube, as shown in Table \ref{table:physical_properties}.  The surface tension coefficient $\gamma$ of the soap water can be estimated using a simplified version of the pendant drop method \cite{surface_tension_drop}. This method utilizes the balance between capillary and gravitational forces, given by $\gamma =\rho g( 6V_{d}/\pi )^{2/3}$ where $\rho$ is the density of soap water and $V_d$ represents the volume of the soap water droplet. 
\begin{table}[h]
\centering
\caption{Recipe of soap water used in our experiment}
\begin{tabular}{|c|c|}
\hline
ingredient            & weight \\ \hline
oleic acid soap & 4.77 $\pm 0.01$ g \\ \hline
water           & 100.1 $\pm 1.1$ g \\ \hline
guar gum        & 0.32 $\pm 0.01$ g \\ \hline
\end{tabular}
\label{recipe}
\end{table}

\begin{table}[h]
\centering
\caption{Physical parameters of soap water and film}
\label{table:physical_properties}
\begin{tabular}{|c|c|c|c|c|}
\hline
                   & \begin{tabular}[c]{@{}c@{}}Density \\ (g/ml)\end{tabular} & \begin{tabular}[c]{@{}c@{}}Viscosity \\ (mPa s)\end{tabular} & \begin{tabular}[c]{@{}c@{}}Surface \\ Tension (N/m)\end{tabular} & \begin{tabular}[c]{@{}c@{}}Thickness \\ ($\mu$m)\end{tabular} \\ \hline
Average            & 0.98                                                     & 60.8                                                        & 0.041                                                              & 5.8                                                            \\ \hline
Standard Deviation & 0.01                                                     & 2.8                                                         & 0.003                                                              & 1.1                                                            \\ \hline
\end{tabular}
\end{table}

The dominant term in the collapse is determined by several dimensionless numbers. Firstly, the Reynolds number Re = $\rho \bar{v_{c}} \bar{h} / \eta \approx 10^{2}$ reveals that the shear viscous stress is smaller than the inertia stress, where  $\bar{v_{c}}$ is magnitude for the time average of  
$dh_{\rm{min}}/d\tau$ $\sim 1.5$ m/s, $\bar{h}$ is the time average of   $h_{\rm{min}}$ $\sim$ 10 mm in roll-off regime, and $\eta$ is the shear viscosity of soap water.  In the meantime, the Bond number Bo = $\rho g R \delta / \gamma \approx 10^{-2}$ and the Weber number We = $\rho v_{s}^{2} R / \gamma \approx 10^{-1}$ reveal two other pieces of important information, i.e., the soap bubble can be considered symmetric as the effect of gravity is small, and the stretching process of the film and bubble in the equilibrium regime is quasi-static, where $g$ is the gravitational acceleration, $\delta$ is the thickness of film and bubble, and $v_s$ is the pulling speed.

\section{Experimental results}
\subsection{Equilibrium regime}
\begin{figure}[h!]
    \centering
	\includegraphics[width=8cm]{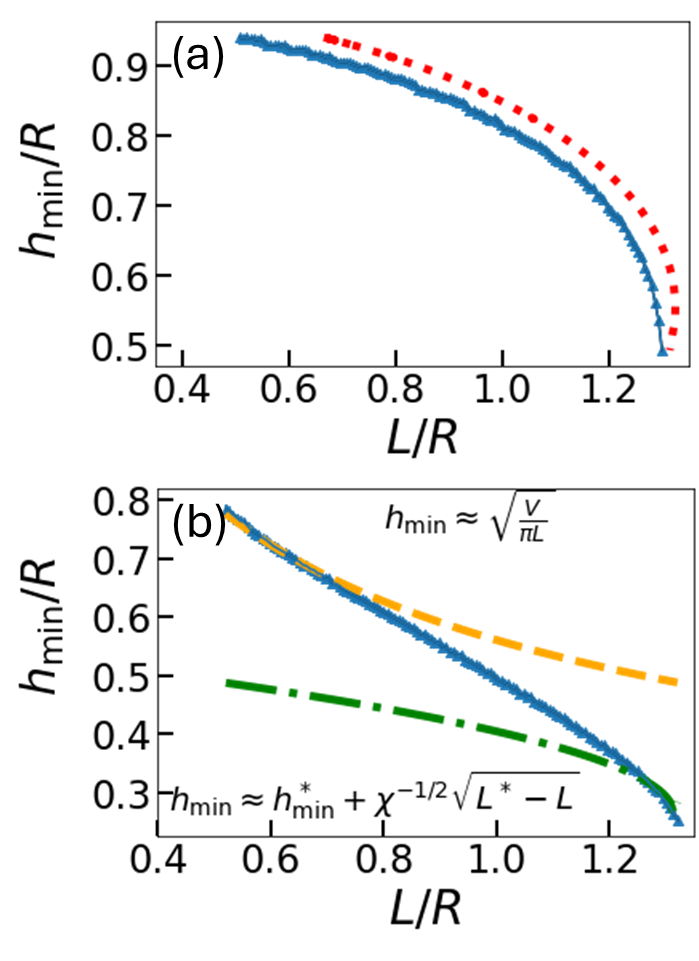}   
    \caption{ Normalized $h_{\rm{min}}/R$ vs. $L/R$ for (a)  film and  (b) bubble during equilibrium regime where $R=$  20 mm, $v_{s}=$  16 mm/s and $V=$ 14 ml. The experimental data are denoted by blue triangles and the error bars are smaller than the symbols.  The red dotted line represents the prediction of a catenoid shape, while the orange dashed and green dash-dotted lines are theoretical predictions. Unnormalized $h_{\rm{min}}$ is plotted as a function of $\tau$ at different $v_{s}$ for (c) film and (d) bubble where $R=$ 11.5 mm and $V=$ 2.9 ml.  (c, d) are rescaled in (e, f) at different $R$ where $v_{s}=$ 16 mm/s and $V/R^3=$ 3.2. The gray and blue background in (c$\sim$f) represents roll-off and equilibrium regimes, while the blue dashed lines fit with the specified functions. }
    \label{fig:sp_volume}
\end{figure}

The profile of the film and bubble within the equilibrium regime is governed by $L$. The impact of the volume conservation constraint is evident in the correlation between $h_{\rm{min}}$ and $L$, as depicted in Fig. \ref{fig:sp_volume}(a, b).  In comparison to the film, the bubble roughly conserves its air volume when pulled apart, as certified in Appendix C. The surface of the bubble undergoes a transition from convex to concave, whereas a film maintains a convex shape throughout the equilibrium regime. This distinction will be elaborated in the theory section to explain the positive second derivative of $h_{\rm{min}}(L/R)$ observed in bubbles, in contrast to the negative one in films \cite{Chen_steen_numerical, robinson_steen_exp,cryer_steen_exp}, as illustrated in Fig. \ref{fig:sp_volume}(a, b).

\subsection{Breakup regime}
\begin{figure}[h!]
\centering
	\includegraphics[width=8cm]{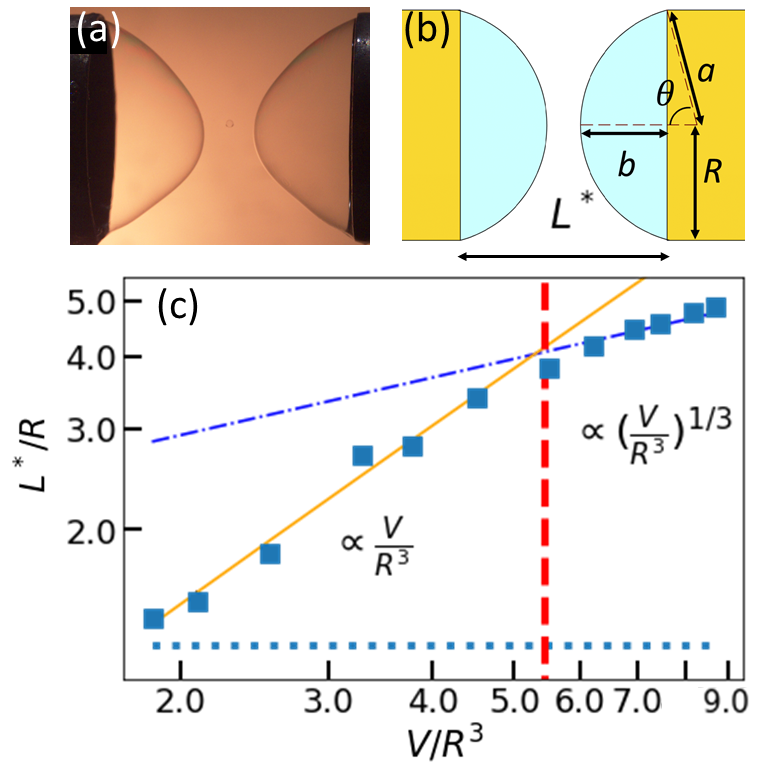}
	
    \caption{ (a) Bubble was split in half after pinch-off. (b) Schematics for the parameters $b$ and $L^*$. (c) Blue squares show the experimental results in a full-log plot for dimensionless critical length ${L^{*}}/{R}$ vs. volume ${V}/{R^{3}}$. The red dashed line ${V}/{R^{3}}=5.44$ separates two regions with different exponents, 1 and 1/3 for the orange and purple dash-dotted lines and the blue dotted line indicates the value of $L^*$ for the film.}	
    \label{fig:lvp3}   
\end{figure}

Unlike the flat surfaces observed on rings in the case of films, two spherical bubbles survive the breakage and form on the caps in the case of bubbles. The spherical shape is preferred to minimize surface energy, and the corresponding height $b$, as defined in Fig.  \ref{fig:lvp3}(a, b), can be calculated accordingly. Simultaneously, the critical length $L^{*}$ at which irreversible processes are initiated can be theoretically determined by considering the breakdown of the solution and minimizing the potential energy in the equilibrium regime. By comparing these two lengths, we observe that $L^{*}$ is not only greater than $2b$, which explains the necessity for both bubbles to retract and eventually break, but it is also roughly equal to $5b/2$. This observation is validated in Appendix D. Depending on the value of $V$, we anticipate two scenarios for the remaining bubble. Through straightforward calculations in Sec. IV C, it is found that $L^{*}/R\propto (V/R^{3})^\beta$ with $\beta = 1$ for $V/R^{3} \ll 5.44$ and $\beta = 1/3$ otherwise. This prediction of $L^*$ vs. $V$  is effectively confirmed by the results shown in Fig.  \ref{fig:lvp3}(c).
\section{Theoretical derivations}
Theoretical analysis of the global geometric features of a collapsing soap film has been performed in detail by Goldstein {\it et al.} \cite{PhysRevE.104.035105}, including the fascinating stages near pinch-off. They showed that the underlying modes are qualitatively the same for mean curvature flow and Euler flow, although the dynamics of their underlying modes are very different. The mechanism that led to pattern formation beyond the critical catenoid was also explained. In contrast, a soap bubble is considerably more difficult to tackle by theoretical means due to the long-ranged mediation of pressure. Notwithstanding, we still managed to derive analytic expressions for some properties primarily in the equilibrium regime after making reasonable approximations. 

\subsection{ Equilibrium regime}
\begin{figure}[h!]
	\centering
	\includegraphics[width=6cm]{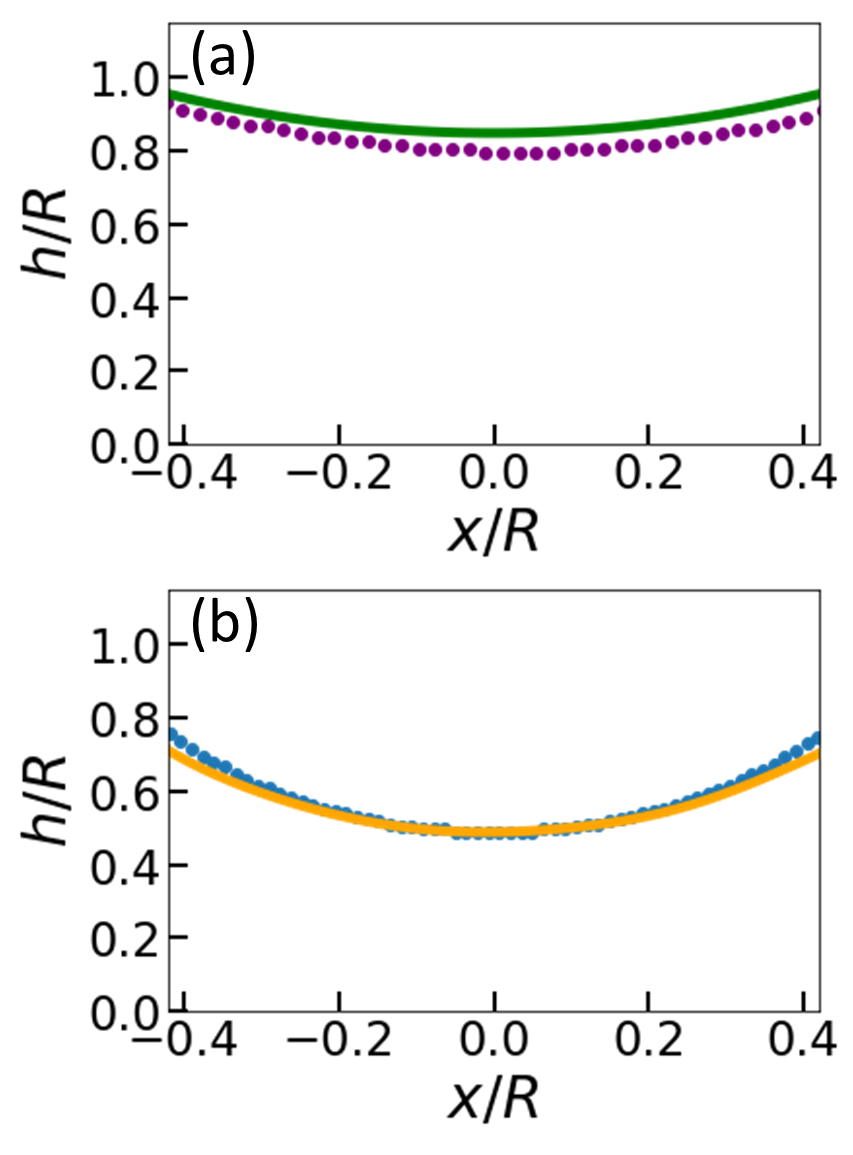}
    \caption{(a) Film and (b) bubble profile in equilibrium regime where $R= 20$ mm, $v_{s}=16$ mm/s, and $V=26$ ml. Green and orange lines are numerical results, while purple and blue dots are experimental data.}
    \label{fig:numerical_exp}
\end{figure}

The profile of the bubble in the equilibrium regime is governed by Eq. (\ref{eq:euler}), which cannot be solved analytically. To employ a numerical method such as finite-difference, we differentiate Eq. (\ref{eq:euler}) with respect to $x$ to obtain
\begin{eqnarray}
h'' = \frac{1+h'^2}{h}+\frac{\lambda}{\gamma}\big(1+h'^2\big)^{3/2}
\label{Eq:fd}
\end{eqnarray} 
In the case of a soap film, the Lagrange multiplier $\lambda$ should be set to zero. The agreement between the numerical and experimental profiles in the equilibrium regime for soap films and bubbles in Fig.   \ref{fig:numerical_exp}(a) indicates that the stretching process can be considered quasi-static and the influence of gravity can be neglected. The small deviations observed in Fig.   \ref{fig:numerical_exp}(b) are attributed to the estimation of $\lambda$.
In our experiments, we can stretch both bubbles and films horizontally or vertically. While measures can be taken to reduce sagging and asymmetrical contours caused by gravity by minimizing the volume $V$, it remains uncertain whether the critical behavior at pinch-off will be affected. Theoretical calculations can help us address this concern and provide analytical expressions for quantities of interest, such as $h_{\rm min}$, while highlighting the influence of volume conservation. To understand how the parameters affect the contour of the bubble, we begin by minimizing the total energy for the bubble:
\begin{eqnarray}
 \frac{U}{2} = \int_{0}^{L/2}\left[\gamma \cdot 2\pi h\sqrt{1+h'^{2}} + \lambda \left(\pi h^{2}-\frac{V}{L} \right)\right]dx
\label{energy}
\end{eqnarray}
where the first integration calculates the surface area, while the second term reinforces the constraint of volume conservation by introducing the Lagrange multiplier $\lambda$.  
Applying the Euler-Lagrange equation \cite{marion} then gives
\begin{eqnarray}
-\gamma\frac{1}{h\sqrt{1+h'^2}}+\gamma\frac{h''}{{(1+h'^2)}^{3/2}}=\lambda
\label{eq:euler1}
\end{eqnarray}
where the product of surface tension and mean curvature on the right-hand side gives the capillary pressure according to the Young-Laplace equation, and $-\lambda$ can be identified as the pressure difference of air across the soap membrane. Note that the factor of $1/\sqrt{1+h'^2}$ comes from the projection onto the normal direction of the surface. 

Since the integrand of Eq. (\ref{energy}) has no explicit dependence on $x$, it is simpler to appeal to the second form of the Euler-Lagrange equation \cite{marion} to obtain
\begin{eqnarray}
\frac{h}{\sqrt{1+h'^{2}}}+\frac{\lambda}{2\gamma}h^{2} = h_{\rm {min}}+ \frac{\lambda}{2\gamma}h_{\rm{min}}^{2}. 
\label{eq:euler}
\end{eqnarray}
After some approximations and transportation detailed in Section 6.2, we derive 
\begin{eqnarray}
\frac{L}{2h_{\rm{min}}} &\approx&  \frac{2\sqrt{\frac{R}{h_{\rm{min}}}-1}-\frac{k+4 \xi}{3}\Big(\sqrt{\frac{R}{h_{\rm{min}}}-1}\Big)^{3}}{\sqrt{2+4 \xi}}
\label{length_app}
\end{eqnarray}
and 
\begin{eqnarray}
\frac{V}{2 \pi h_{\rm{min}}^3} \approx \frac{2\sqrt{\frac{R}{h_{\rm{min}}}-1}-\frac{k+4 \xi-4}{3}\Big(\sqrt{\frac{R}{h_{\rm{min}}}-1}\Big)^{3}}{\sqrt{2+4 \xi}}
\label{volume_app}
\end{eqnarray} 
where $\xi \equiv \frac{\lambda}{2\gamma}h_{\rm{min}}$ and $k=(1+2\xi-4\xi^2)/(2+4\xi)$.
By comparing the leading-order term, we obtain
\begin{eqnarray}
h_{\rm min} \approx \sqrt{\frac{V}{\pi L}}
\label{cylinder}
\end{eqnarray}
which matches the data in early equilibrium regime for Fig.  ~\ref{fig:sp_volume}(b). 

\begin{figure}[h!]
	\centering
	\includegraphics[width=6cm]{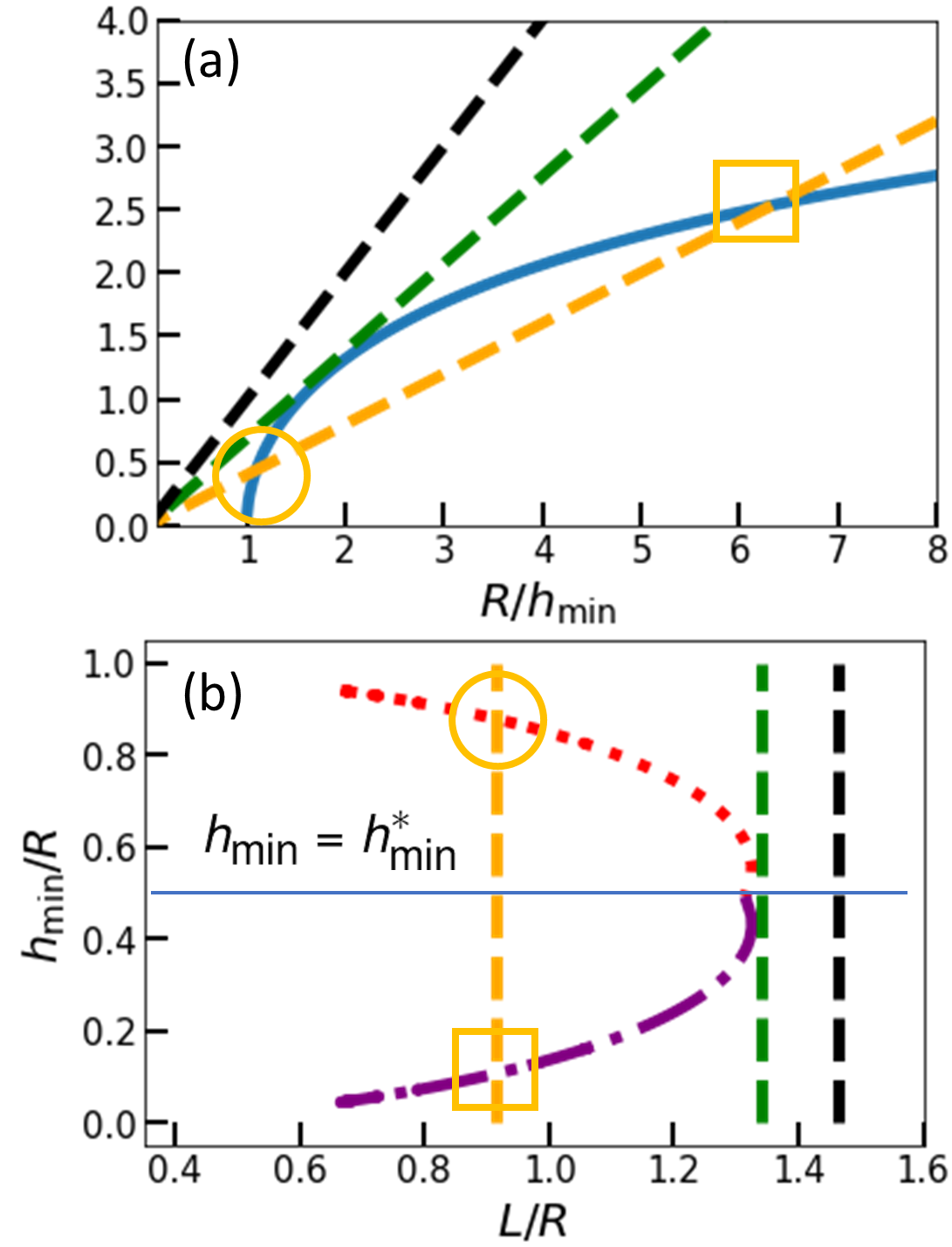}
	
    \caption{(a) The right- and left-hand sides of  Eq. (\ref{length_f}) with $\xi =0$ are plotted in solid and dashed lines as a function of $R/h_{\rm{min}}$.   Three scenarios are possible by increasing the slope $L/(2R)$ with zero, one, and two interceptions that are denoted respectively by the black, green, and yellow dashed lines.  The green line defines the critical $L^*$ and $h^*_{\rm{min}}$. In the meantime, since we expect $h_{\rm min}$ to shrink as $L$ lengthens, the solution highlighted by the yellow square should be discarded.  This unphysical solution is further represented by the purple dash-dot line in (b) that shows $h_{\rm{min}}/R$ vs. $L/R$. }
    \label{fig:eq_op_reverse}

\end{figure}

The derivations from Eq. (\ref{eq:euler}) to Eq. (\ref{volume_app}) illustrate the crucial role of the Lagrange multiplier $\lambda$ and how it mathematically prevents the bubble from adopting the catenoid profile observed in films. These derivations also emphasize the significance of $\lambda$ and the subsequent parameter $\xi$. Without them, the condition $L/(2h_{\rm min}) \approx \sqrt{2(R/h_{\rm{min}}-1}) \ll 1$ would hold, implying that the cylindrical shape is only possible when the two rings are very close together in films. For bubbles, an additional condition stated in Eq. (\ref{volume_app}) requires the denominator to be comparable to $\sqrt{R/h_{\rm{min}}-1}$. This renders a finite value of $L$. 
We can draw insights from the analysis of films. By setting $\xi=0$, we plot both sides of Eq. (\ref{length_f}) as a function of $R/h_{\rm{min}}$ in Fig.   \ref{fig:eq_op_reverse}(a).

When $L>L^*$, there is no intersection, indicating a problem with the starting point for minimizing the surface energy. This aligns with our expectation that films will naturally collapse at large $L$, requiring the minimization of action instead. When $L=L^*$, both sides of the equation become tangent. The same is expected for bubbles, meaning that we should differentiate both sides of Eq. (\ref{length_f}) with respect to $h_{\rm{min}}$ to determine the location of $L^*$. There are two intersections for $L<L^*$, but one of them is unphysical, as explained in Fig.   \ref{fig:eq_op_reverse}(a). When we plot $h_{\rm{min}}$ against $L$, the solution should be double-valued until $L$ reaches $L^*$, as shown in Fig.   \ref{fig:eq_op_reverse}(b). Therefore, we can approximate $L(h_{\rm{min}})\approx -\chi (h_{\rm{min}}-h^*_{\rm{min}})^2 +L^*$, where $\chi$ is a constant, for values of $h_{\rm{min}}$ close to the critical neck radius $h^*_{\rm{min}}$, beyond which the neck collapses spontaneously. Simple rearrangement gives 
\begin{eqnarray}
h_{\rm{min}} \approx h^*_{\rm{min}}+ \chi^{-1/2} \sqrt{L^{*}-L}.
\label{bubble_equillbrium_end}
\end{eqnarray}
Our confidence in Eqs. (\ref{cylinder}) and (\ref{bubble_equillbrium_end}) is supported by their accurate prediction of an inflection point in Fig.   \ref{fig:sp_volume}(b), which arises from the fact that the curvatures have opposite signs.

\subsection{Breakup regime}
The collapse speed of the neck is extremely rapid. As we will discuss later, there are two additional complexities before the final breakage occurs. First, in the pinch-off regime, two necks will form. Second, this is followed by the breaking stage, during which a satellite bubble is created in the middle of these two necks after the hollow, thin tube connecting them transforms into a liquid string. There is no gas leakage during the breaking and relaxation stages, and the volume of the satellite bubble can be neglected. Therefore, $V$ should be equal to the combined volume of the two remaining bubbles after breakage. We can directly establish the relationship between $V$ and $L^{*}$, which can be estimated from Fig.   \ref{fig:lvp3}(b) where the chord length equals $2R$. Using the geometry shown in Fig.   \ref{fig:lvp3}(b), we can obtain the volume of each partial sphere 
\begin{eqnarray}
\frac{V}{2}= \frac{\pi b}{6}\left(3R^{2}+b^2\right)
\label{volume_dome}
\end{eqnarray}
where $b\approx L^*/2$.  It is easy to show that the first term dominates when ${L^{*}}/{R} \ll 2 \sqrt{3}$ and
\begin{eqnarray}
\frac{V}{R^{3}} \cong \frac{\pi}{2}\frac{L^{*}}{R}
\label{eq:small V},
\end{eqnarray}
while the second term prevails to give
\begin{eqnarray}
\frac{V}{R^{3}} \cong \frac{\pi}{24} \left(\frac{L^{*}}{R} \right)^{3}
\label{eq:big V}
\end{eqnarray}
at the other extreme. The predictions of Eqs. (\ref{eq:small V}) and (\ref{eq:big V}) are vindicated by Fig.   \ref{fig:lvp3}(c).

\section{Conclusion}
This study combines experimental observations with theoretical analysis to elucidate the shape evolution of soap bubbles under the constraint of volume conservation. 
Since soap films and bubbles are short-lived, previous scientists chose to stretch them vertically because they were easier to burst when placed horizontally. This arrangement rendered the asymmetry in shape due to gravity. In this work, we managed to overcome this technical difficulty by adding guar gum. 
With the special recipe in Table I, the lifetime can be lengthened by five folds to enable the experimental observations of horizontal stretching, effectively minimizing the effects of gravity. 

By contrasting soap bubbles with open films, we demonstrate the critical role of volume conservation in influencing surface tension-driven dynamics. 
Mainly, the shape of bubbles exhibits a convex-to-concave transition, in contrast to being always convex for films. 
Theoretically, this is shown to be linked to the Lagrange multiplier associated with the volume conservation.
Although not included in this work, preliminary studies of ours showed that the effect of such a constraint persists in influencing the behavior of non-equilibrium fluid systems and underscoring universal behaviors in surface shape.
Potential applications involving the volume conservation and breakup of droplets span from the design of microfluidic devices to the study of biological membranes, offering a foundation for exploring broader implications of fluid dynamics and surface tension in physical and applied contexts.

\section*{acknowledgment}

We are grateful to C. Y. Lai and J. R. Huang for useful discussions and thank P. Yang and J. C. Tsai for the use of high-speed cameras. Financial support from the Ministry of Science and Technology in Taiwan under Grant No. 111-2112-M007-025 and No. 112-2112-M007-015 is acknowledged.

\appendix

\section{Asymmetrical breakup}

Occasionally it was observed that the two partial spheres could be of a different 
size. We believe it was caused by the additional liquid that pends at the bottom 
of the bubble. When this happens, $L^*$ will become smaller than expected from
Eq. (\ref{eq:big V}). The following calculation can verify this anomaly. First, we distinguish the left from the right spheres by appending subscripts $L$ and $R$. Second, 
differentiating the volume
\begin{eqnarray}
dV = 0 = \frac{dV_{L}}{dL^*_{L}}dL^*_{L}+\frac{dV_{R}}{dL^*_{R}} dL^*_{R}
\label{eq:lv_lf}
\end{eqnarray}
where we have separated $L^*$ into three segments - $L_L^*$, the gap, and $L_R^*$. Neglecting 
the small gap, we have $dL = dL_L^* + dL_R^*$. Plugging this into Eq. (\ref{eq:lv_lf}) gives
\begin{eqnarray}
\frac{dV_R}{dL^*_{R}} - \frac{dV_L}{dL^*_{L}} dL^*_{L} = \frac{dV_R}{dL^*_{R}} dL^*
\label{eq:422}
\end{eqnarray}
where the right parenthesis is equivalent to $\frac{d^2 V_R}{d(L_R^*)^2}$ times $L_R^* - L_L^*$. The
former is positive definite from Eq. (\ref{eq:lv_lf}), and so Eq. (\ref{eq:422}) requires $dL^*$ to share 
the same sign of $dL_L^*$ if $L_R^* > L_L^*$. A short summary of this cute derivation: the
bigger the size difference between spheres, the shorter $L^*$ is.

\section{Detailed steps of derivation for Section IV.A}
After some transpositions, Eq. (\ref{eq:euler}) becomes
\begin{eqnarray}
h'=\sqrt{\left(\frac{h}{h_{\rm{min}}+\frac{\lambda}{2\gamma}h_{\rm{min}}^{2}-\frac{\lambda}{\gamma}h^{2}} \right)^{2}-1}.
\label{eq:first_d}
\end{eqnarray}
Solving this differential equation will enable us to obtain information on the contour $h(x)$:
\begin{eqnarray}
x=\bigints_{h_{\rm{min}}}^{h} dh/\sqrt{\left(\frac{h}{h_{\rm{min}}+\frac{\lambda}{2\gamma}h_{\rm{min}}^{2}-\frac{\lambda}{\gamma}h^{2}} \right)^{2}-1}
\label{x}
\end{eqnarray}
where $0\le x \le L/2$. There is no need for an additional constant in Eq. (\ref{x}) since $h=h_{\rm min}$ occurs at $x=0$.

By implementing the boundary condition that $h(L/2)=R$ and volume conservation, we get
\begin{eqnarray}
\frac{L}{2h_{\rm{min}}} = \bigints_{1}^{\frac{R}{h_{\rm{min}}}}{dy}/{\sqrt{\Big(\frac{y}{1+ \big(1-y^{2}\big)\xi}\Big)^{2}-1}}
\label{length_f}
\end{eqnarray} 
and
\begin{eqnarray}
\frac{V}{2\pi h_{\rm{min}}^{3}} = \bigints_{1}^{\frac{R}{h_{\rm{min}}}}{ y^{2}dy}/{\sqrt{\Big(\frac{y}{1+ \big(1-y^{2}\big)\xi}\Big)^{2}-1}}
\label{volume_f}
\end{eqnarray}
where a change of variable $y=h/h_{\rm{min}}$ has been performed to render the parameters dimensionless and $\xi \equiv \frac{\lambda}{2\gamma}h_{\rm{min}}$.
By setting $\lambda =0$, Eq. (\ref{length_f}) will revert to depicting a film and give us $h_{\rm{min}}/R$ vs. $L/R$ in agreement with Fig.  ~\ref{fig:sp_volume}(a). 
Setting $u=y-1$ rewrites the right-hand-side of Eq. (\ref{length_f}) and (\ref{volume_f}) as
\begin{eqnarray}
\bigints_{0}^{\frac{R}{h_{\rm{min}}}-1}\frac{[1-(2u+u^2)\xi]du}{\sqrt{(u+1)^2-[1-(2u+u^2)\xi]^2}}
\label{length_u}
\end{eqnarray}
and
\begin{eqnarray}
\bigints_{0}^{\frac{R}{h_{\rm{min}}}-1} \frac{[1-(2u+u^2)\xi](1+2u+u^2)du}{\sqrt{(u+1)^2-[1-(2u+u^2)\xi]^2}}.
\label{volume_u}
\end{eqnarray}
During the stretching of the bubble, there is a certain period where $h_{\rm{min}}$ is close to $R$ and $0<u<R/h_{\rm{min}}-1 \ll 1$. Expanding the expression to $\mathcal{O}(u^2)$, the denominator can be simplified as follows:

\begin{equation}
    \begin{aligned}
        \sqrt{(u+1)^2-[1-(2u+u^2)\xi]^2} \\
        \approx \sqrt{(2+4 \xi )u+(1+2 \xi- 4 \xi^2)u^2}.
    \end{aligned}
    \label{denominator}
\end{equation}

With further rearrangement
\begin{eqnarray}
\frac{1}{\sqrt{(2+4 \xi )u+(1+2 \xi- 4 \xi^2)u^2}} \approx \frac{1-\frac{ku}{2}+\frac{3k^2u^2}{8}}{\sqrt{(2+4 \xi)u}}
\label{taylor}
\end{eqnarray}
where $k=(1+2\xi-4\xi^2)/(2+4\xi)$. Neglecting terms higher than $\mathcal{O}(u^2)$ allows us to analytically solve the integration of Eqs. (\ref{length_u}) and (\ref{volume_u}) and expand them as polynomials of the small number $\big(R/h_{\rm{min}}-1 \big)$.

\section{Is the volume of bubbles truly conserved?}
Young-Laplace equation, $\Delta P = \gamma(1/r_1+1/r_2)$, allows us to determine the radius of curvature for the surface by the pressure difference. By using the characteristic length $h^{*}$, we estimated that $\Delta P / P_{1} \approx 10^{-3}$ where $P_{1}=1$ atm is the initial pressure of air inside the bubble. Treating the air as being ideal, we can employ the equation of state  $PV =Nk_B T$ where the particle number $N$ and temperature $T$ are fixed, and $k_B$ is the Boltzmann constant. Because the pressure is inversely proportional to the volume, the ratio of $\Delta V$ to  $V_{1}$ is $10^{-3}$ which 
indicates that $\Delta V$ is negligible.
The small change in volume will become significant when the sum of volume for cap A and tube, roughly 17.0 and 41.5 ml, is $10^{3}$ times the size of a soap bubble. These two volumes are of the same order in our experiment and therefore we do not need to worry about the extreme case.
\section{Evidence for $b \approx 0.4L^*$}

\begin{figure}[h]
	\centering
	\includegraphics[width=6cm]{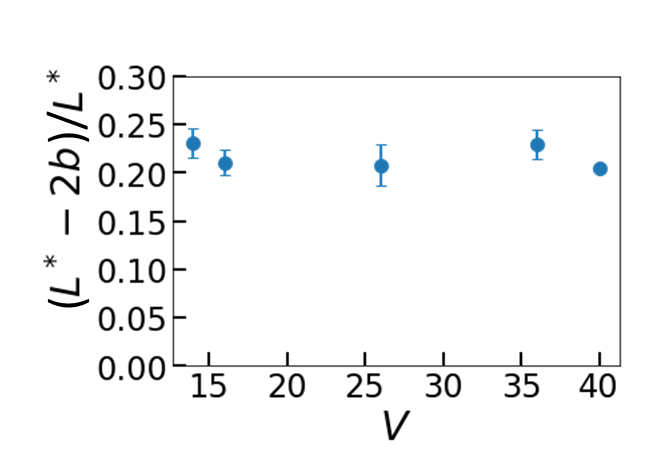}
	
    \caption{Data for $(L^*-2b)/L^*$ vs. $V$ where $R= 20$ mm and $v_{s}=16$ mm/s.}
    \label{fig:lv_check}
\end{figure}

According to our theoretical calculation in  Section IV.B of the main text, the gap width equals $L^*-2b$ between the tips of remnant soap bubbles after breakup. The relation $b = 2L^*/5$ comes from the experimental result $(L^*-2b)/L^* \sim 0.2$, as shown in Fig. \ref{fig:lv_check}.

\nocite{*}
\bibliography{aipsamp}

\end{document}